\magnification= \magstep 1
\global\newcount\meqno
\def\eqn#1#2{\xdef#1{(\secsym\the\meqno)}
\global\advance\meqno by1$$#2\eqno#1$$}
%
\global\newcount\refno
\def\ref#1{\xdef#1{[\the\refno]}
\global\advance\refno by1#1}
\global\refno = 1
%
\hsize = 17.5 true cm
\vsize = 24 true cm
\tolerance 10000
%
\def\s#1{{\bf#1}}
\def\tr{{\rm tr}}
\def\ln{{\rm ln}}
\def\hf{{1\over2}}
\def\dt{\Delta t}
\def\dx{\s x-\s y}
\def\bx{\partial^2}
\def\xy{\s x+\s y}
\def\ur{\tilde U}
\def\urv{\tilde V}

\baselineskip 12pt plus 1pt minus 1pt
\vskip 1.5cm
\centerline{\bf RENORMALIZATION GROUP IN QUANTUM MECHANICS}
\vskip 1cm
\centerline{Janos Polonyi}
\vskip .5cm
\centerline{\it Laboratory of Theoretical Physics}
\centerline{\it Louis Pasteur University}
\centerline{\it 3 rue de l'Universit\'e 67084\ \ Strasbourg\ \ Cedex\ \ France}
\vskip .4cm
\centerline{\it Department of Atomic Physics}
\centerline{\it Lorand E\"otv\"os University}
\centerline{\it Puskin u. 5-7  \ \ 1088 \ \ Budapest \ \ Hungary}
\vskip 3cm
\centerline{Submitted to Annals of Physics}
\vskip 1cm
\centerline{{\bf ABSTRACT}}
The running coupling constants are introduced in Quantum
Mechanics and their evolution is described by the help of the 
renormalization group equation. The harmonic oscillator and the propagation
on curved spaces are presented as examples. The hamiltonian and the
lagrangian scaling relations are obtained. These evolution equations are
used to construct low energy effective models.
\medskip
\vfill
\eject
\centerline{\bf 1. INTRODUCTION}
\medskip
\nobreak
\xdef\secsym{1.}\global\meqno = 1
\medskip
The renormalization group provides us with the systematic description of the 
dependence of the fundamental laws and constants on the observational
scale \ref\wilsrgr. This scale dependence is rather involved as we
pass from the realm of one interaction to another or we are in
the vicinity of a phase transition. The scaling laws found in the 
solutions of the renormalization group equations reflect the layered
structure in which different degrees of freedoms contribute to the dynamics.

Thus one might think that the application of the renormalization group 
method is not too illuminating for a simple system, in particular a
single particle in nonrelativistic Quantum Mechanics. This may not be so.
Compare the path integral of a 0+1 dimensional Quantum 
Field Theory for a three component field, $\phi^k(x)$, $k=1,2,3$,
\eqn\qftpi{\int D[\phi]e^{{i\over\hbar}\int dx
[\hf({\partial\phi^k\over\partial x})^2-V(\phi^j)]},}
with the path integral expression for a 3 dimensional Quantum Mechanical
system given by the hamiltonian $H={\s p^2\over2m}+V(\s x)$, 
\eqn\pathd{\eqalign{<\s x_f|e^{-{i\over\hbar}tH}|\s x_i>
&={\rm lim}_{N\to\infty}
\biggl({m\over2\pi i\hbar\dt}\biggr)^{{3\over2}(N+1)}
\prod_{j=1}^N\int d\s x_je^{{i\over\hbar}\dt\sum_{j=0}^N
{m\over2}({\s x_{j+1}-\s x_j\over\dt})^2-V(\s x_j)}\cr
&=\int D[\s x]e^{{i\over\hbar}\int dt'[({d\s x\over dt})^2-V(\s x)]},}}
where $\dt={t\over N}$ $\s x_0=\s x_i$ and $\s x_{N+1}=\s x_f$.
The similarity of \qftpi\ and \pathd\ is the formal reason for the
emergence of ultraviolet divergences, running coupling constants
and other ingredients of the renormalization group in Quantum Mechanics.
The present paper contains some simple observations concerning the 
application of the renormalization group in Quantum Mechanics.

The renormalization group usually appears in different contexts. It can be used
(i) to display the scale dependence of the interactions or 
(ii) to eliminate unimportant variables or (iii) to describe the short distance
behavior of products of field operators. Though the mathematical expressions
are usually similar they can be interpreted in different manner. We use
the renormalization group in the sense (i) only.

There have already been some works addressing the renormalization group
in the context of Quantum Mechanics. In Refs. \ref\gosdr, \ref\gupr, 
\ref\manuelr\ and \ref\riverr\
the ultraviolet divergences generated by certain short range singular
potentials are treated in a manner which is similar to Quantum Field Theory.
The goal of the present work is different. Instead of tracing down the 
singular effects of a singular potential we investigate the impact
of the ultraviolet singular `quantum noise' in regular systems.

We start with a Quantum Mechanical particle in the presence of an
external magnetic field,
\eqn\hammd{H={(\s p+\s A(\s x))^2\over2m}+V(\s x).}
The corresponding transition amplitude can be written as \ref\nieuwr,
\eqn\pathmd{\eqalign{<\s x_f|e^{-{i\over\hbar}tH}|\s x_i>
&={\rm lim}_{N\to\infty}
\biggl({m\over2\pi i\hbar\dt}\biggr)^{{3\over2}(N+1)}
\prod_{j=1}^N\int d\s x_j\cr
&e^{{i\over\hbar}\dt\sum_{j=0}^N[
{m\over2}({\s x_{j+1}-\s x_j\over\dt})^2-V(\s x_j)
+{\s x_{j+1}-\s x_j\over\dt}\s A(\hf(\s x_{j+1}+\s x_j))]}.}}
A new problem arises in the presence of a velocity dependent interaction,
namely the operator ordering problem which renders different classical actions
to the same quantum hamiltonian. This problem can be solved formally,
it has been shown that the construction of the quantum
lagrangian of the path integral formalism from the quantum hamiltonian
is well defined and unique when the mid-point prescription is used for
the velocity couplings \ref\orderpr. 

We believe that there is more about the complications of \pathmd\ than the 
ordering ambiguity of quantum mechanics and it is closely related to
the nowhere differentiable nature of the trajectories which saturate 
the path integral. This feature serves as a starting point to motivate 
the use of the renormalization group in Quantum Mechanics.
The hand-waving argument to make the non-differentiability of the trajectories 
plausible for a free particle,
\eqn\pathfd{\eqalign{<\s x_f|e^{{it\over2\hbar}
{\partial^2\over\partial\s x^2}}|\s x_i>={\rm lim}_{N\to\infty}
\biggl({m\over2\pi i\hbar\dt}\biggr)^{{3\over2}(N+1)}
\prod_{j=1}^N\int d\s x_je^{{im\over2\hbar\dt}\sum_{j=0}^N
\Delta_j\s x^2},}}
where $\Delta_j\s x=\s x_{j+1}-\s x_j$ is the following. For the typical 
trajectories each contribution in the exponent is $O(1)$. One the one hand, 
for ${m\Delta\s x^2\over\hbar\dt}>>1$, the fast oscillation suppresses the
contributions in the vicinity of the path.
On the other hand, when ${m\Delta\s x^2\over\hbar\dt}<<1$ then the phase
space of the trajectory is too restrictive, its `entropy' is small.
Actually, the Gaussian integration gives
\eqn\brownian{<\biggl({\Delta x\over\dt}\biggr)^2>={i\hbar\over m\dt},}
after ignoring the end point dependence. The short time 
behavior is clearly dominated by the kinetic energy so our result remains
valid asymptotically in the presence of a potential, too. Note that
\brownian\ is imaginary because the weight of the paths is complex for
real time.

This property of the trajectories is in conflict with our
intuition which is based on classical mechanics. But it reflects an essential
element of Quantum Mechanics, the canonical commutation relations 
\ref\schulmanr. In fact, to verify the matrix elements of the commutator,
\eqn\cancom{\eqalign{<[x,p]>&={m\over\dt}<x_{k+1}(x_{k+1}-x_k)
-(x_{k+1}-x_k)x_k>\cr
&={m\over\dt}<(x_{k+1}-x_k)^2>\cr
&={m\over\dt}{i\hbar\dt\over m}=i\hbar,}}
we have to use \brownian.
The quantum motion is rather disordered for short time observers.
One finds a sort of uncertainty relation,
\eqn\uncert{\Delta v\sqrt{\dt}=\sqrt{\hbar\over m},}
reflecting the fast spreading of the wave-packets for short time.
We may interpret \uncert\ as the indication that the Hausdorff dimension
of the quantum trajectory is twice of the classical one. There is naturally
the question whether the individual trajectories have any significance
for the expectation values. The singular fluctuations may not produce any 
particular scale dependence for the observables. Our strategy to deal with
this question is the introduction of the running coupling constants and
the investigation of their evolution.

Traditionally, the running coupling constants appear in Quantum Field
Theory due to the ultraviolet divergences. The evolution equations
for these coupling constants obtained in the framework of the perturbation
expansion show clearly the manner these divergences build up as the
ultraviolet modes are integrated over in the path integral. 
We shall use this method to identify the effects of the singular structure
of the trajectories. 

It is obvious in classical mechanics that the velocity dependent interactions 
play more important role in short time processes where the initial and final 
conditions are given. What we is expressed in \uncert\ is an additional 
increase of the strength of these interactions at high energy which 
points beyond the semiclassical behavior.

The organization of this paper is the following. The perturbation expansion
and the power counting scheme is worked out in Sections 2 and 3. The
running coupling constants are introduced in Section 4 and the harmonic
oscillator is discussed as a simple example. Section 5 contains
the perturbative one-loop computation of the decimation, the blocking
relation corresponding to the change $\dt\to2\dt$. The infinitesimal form of 
the renormalization group equation and the hamiltonian is worked out 
in Section 6. Low energy effective theories are the subject of Section 7.
Our formalism is applied for systems on curved space in Section 8.
Section 9 discusses the anomalies and the operator ordering problem
from the point of view of the renormalization group. Finally Section 10 is 
reserved for the summary and some open questions.
\bigskip
\centerline{\bf 2. PERTURBATION EXPANSION} 
\medskip
\nobreak
\xdef\secsym{2.}\global\meqno = 1
\medskip
In this Section we briefly review the perturbation 
expansion for \pathmd\ from the point of view of a Quantum Field Theory 
in 0+1 dimension. The quantity
of central importance is the trace of the time evolution operator in the path
integral formalism,
\eqn\pertpth{\eqalign{\tr e^{-{i\over\hbar}TH}&={\rm lim}_{N\to\infty}
\biggl({m\over2\pi i\hbar\dt}\biggr)^{{3\over2}(N+1)}
\prod_{j=0}^N\int d\s x_j\cr
&e^{{i\over\hbar}\dt\sum_{n=0}^N
[{m\over2}({\s x_{n+1}-\s x_n\over\dt})^2-V(\s x_n)
+{\s x_{n+1}-\s x_n\over\dt}\s A(\eta_+\s x_{n+1}+\eta_-\s x_n)]},}}
where $\s x_0=\s x_{N+1}$. The left hand side of \pertpth\ may have 
ultraviolet and infrared divergences. To separate the overall 
ultraviolet divergence from the Green functions we use 
$m\to m+i\epsilon$ with infinitesimal $\epsilon$. The infrared
divergence may arise for interactive systems with continuous energy
spectrum. To regulate this divergence we place the system into a large but 
finite quantization box. Our primary interest in the rest of the paper
is the ultraviolet behavior and we shall not attempt to remove the infrared
cut-off. 

The vector potential is evaluated at the
intermediate point which is not necessarily the midpoint,
$\eta_\pm=\hf\pm\eta\not=\hf$. The potentials are written in power series,
$V(\s x)=V+V^ix^i+V^{ij}x^ix^j+V^{ijk}x^ix^jx^k\cdot$, and 
$A_\ell(\s x)=a_\ell+a_\ell^ix^i+a_\ell^{ij}x^ix^j+a_\ell^{ijk}x^ix^jx^k+\cdots$
and the perturbation expansion of the path integral consists of expanding
the exponential in the right hand side of \pertpth\ in the 
coupling constants $V$ and $a$. 
The contributions can be labeled by Feynman graphs just as in Quantum Field
Theory.

Note that the coupling constants of the potential can be
determined from the `scattering amplitudes' defined in analogy with
Quantum Field Theory. In fact, let us denote the state with the
wave function $x^{j_1}\cdots x^{j_n}\psi_0(\s x)$, where $\psi_0(\s x)$ 
corresponds to the ground state by $|x^{j_1}\cdots x^{j_n}>$. Then the 
amplitude,
${\rm lim}_{t\to\infty}<x^{j_1}\cdots x^{j_n}|U_i(t)|x^{k_1}\cdots x^{k_m}>$,
where $U_i(t)$ is the time evolution operator in the interaction picture
can be written as the sum of Feynman graphs with $n+m$ external legs.
On the one hand, the Moller operator can be reconstructed from the 
transition amplitudes.
On the other hand, starting from the Moller operator 
one can generate these amplitudes. 

The Fourier expanded form of the trajectories is
\eqn\trf{\s x(\dt n)={1\over N}\sum_{k=1}^Ne^{i{2\pi\over N}kn}\s X_k,}
which becomes
\eqn\ctrf{\s x(t)=\int_{-\Lambda/2}^{\Lambda/2}{d\omega\over2\pi}
e^{i\omega t}\s X(\omega),}
with $\s X(k{2\pi\over\dt})=\dt\s X_k$ and
$\Lambda={2\pi\over\dt}$ when the I. R. cut-off is removed, 
$T=\dt N\to\infty$. The free propagator is given by
\eqn\freepre{G^{ij}_0(\dt(n-m))=
\tr T[e^{-{i\over\hbar}\int dtH_i(t)}x^i(\dt n)x^j(\dt m)]=
\delta^{ij}{i\hbar T\over4mN^2}\sum_{k=1}^N
{e^{i{2\pi\over N}k(n-m)}\over\sin^2{\pi\over N}k+i\epsilon},}
or
\eqn\freepr{G^{ij}_0(t_1-t_2)=
\delta^{ij}{i\hbar\over m}\int_{-\Lambda/2}^{\Lambda/2}{d\omega\over2\pi}
{e^{i\omega(t_1-t_2)}\over{1\over\dt^2}4\sin^2(\omega\dt/2)+i\epsilon}.}

It is worthwhile noting that this perturbation expansion is essentially 
different from that of the operator formalism. This is reflected in the
propagator $O(\omega^{-2})$ in \freepr\ as opposed to the one with 
$O(\omega^{-1})$ of the operator formalism. The U.V. divergence in the
average of the velocity square,  
\eqn\avvs{{1\over T}<\int_0^Tdt({d\s x\over dt})^2>=
-{1\over T}<\int_0^Tdt\s x{d^2\s x\over dt^2}>,}
comes from the linear divergence
\eqn\ldivpr{{1\over\dt^2}(G^{ij}_0(\dt)+G^{ij}_0(-\dt)-2G^{ij}_0(0))=
-\delta^{ij}{i\hbar\over m}\int_{-\Lambda/2}^{\Lambda/2}{d\omega\over2\pi},}
which is the immediate consequence of the $O(\omega^{-2})$ behavior of the 
propagator. It is interesting to note that the canonical commutation relation 
is lost and the dynamics become classical with propagators
$O(\omega^{-p})$, $p>2$. 

Another interesting consequence of the non-differentiability is the
appearance of ultraviolet divergences in quantum mechanics. 
Consider the general potential
$U(x,\dot x)=u^ix^i+u_i\dot x^i+u^{ij}x^ix^j
+u_{ij}\dot x^i\dot x^j+u_i^j\dot x^ix^j+\cdots$,
in the lagrangian. The dot stands for time derivative and the
midpoint prescription is used for the coordinate argument $x$.
In the perturbative treatment the parameter 
$u^{i_1,\cdots,i_n}_{j_1,\cdots,j_m}$ is the coupling constant for a vertex 
of order $n+m$. The frequencies corresponding to $m$ legs multiply
its contribution. It is easy to see that an internal closed line connecting 
two velocity legs of the same vertex gives a linear divergent contribution
in the graphs. It is this linear divergence, \ldivpr, which is responsible
for several interesting ultraviolet phenomenon discussed 
in the rest of this paper. The systematic classification of the
divergences will be achieved by evoking the power counting method.

We shall use this perturbation expansion
to demonstrate the dependence of the path integral on the choice
of the intermediate points where the vector potential is evaluated.
The leading order $\eta$ dependence of \pertpth\ is
\eqn\linvp{\eqalign{&{i\eta\over\hbar}\dt
\sum_n<{1\over\dt}(x^i_{n+1}-x^i_n)a^{ij}(x^j_{n+1}-x^j_n)>\cr
&=-{i\eta\dt\over\hbar}a^{ij}\int dt{d^2G_0^{ij}(t)\over dt^2}\Big\vert_{t=0}
=-a^{ii}{\eta T\dt\over m}\int_{-\pi/\dt}^{\pi/\dt}{d\omega\over2\pi}\cr
&=-{\eta T\over m}\partial_iA_i(0),}}
indicating that the path integral is not gauge invariant unless $\eta=0$. 
Had the trajectories been differentiable, \ldivpr\ would 
have been finite and the faster decreasing propagators would have made 
\linvp\ vanishing. The dependence of the path integral on $\eta$
could have been guessed at the very beginning by noting that $\eta\not=0$
implies the appearance of the difference of two consecutive coordinate
values, $\s x_{n+1}-\s x_n$, in the first line of \linvp.
This term is multiplied by another factor of 
$\s x_{n+1}-\s x_n$ in the last expression of the exponent 
of \pertpth. Due to \brownian\ this square is proportional with $\dt$
so the contribution is finite and its sum along the trajectory is 
proportional to the total time, $T$. What is important in this simple
argument is that the inherent ultraviolet divergences of the perturbation
expansion may make terms which vanish in the classical continuum limit,
$\dt\to0$, survive the quantum continuum limit.

Note that it is \brownian, the amplification of the $<(\s x_{n+1}-\s x_n)^2>$
compared to the classical motion which requires the replacement of the 
classical calculus in the path integral by the It\^o-calculus \ref\itocalr. 
In the It\^o calculus one has to keep track of 
the terms of the Taylor expansion in one additional order of accuracy 
when the transformation $\s x\to\s x'=\s F(\s x)$ is made. 
The new contribution to the It\^o-integral which
arises from the nonlinear change of variables appears as a time integral 
for $t>\dt$ since $\sum\Delta_jx^2={i\hbar\over m}\sum_j\dt$, see Ref.
\ref\itopr\ for the details.
\bigskip
\centerline{\bf 3. POWER COUNTING AND RENORMALIZABILITY} 
\medskip
\nobreak
\xdef\secsym{3.}\global\meqno = 1
\medskip
One can easily develop the power counting for \pertpth. The path integral
is derived in `lattice regularization', for finite $\dt$. The propagator
has -2 degree of ultraviolet divergence in energy space, according to \freepr. 
Each time derivative corresponds to the multiplication by
${1\over\dt}(e^{i\omega\dt}-1)$, which has 1 degree of ultraviolet divergence
in agreement with the general construction of Ref. \ref\rieszpcr. 
To find the superficial degree of divergence of the loop integrals we can 
follow the usual dimensional argument after separating the contribution of 
the frequency independent Planck constant $\hbar$. 
The dimension of the action is $[S]=ML^2T^{-1}$,
where $M,L$ and $T$ stands for the mass, length and time. Thus
the dimension of the coupling constant $G_{ds}$ in the action
\eqn\actd{S=\dt\sum_{ds}
G_{ds}\biggl({\s x_{n+1}-\s x_n\over\dt}\biggr)^d
\biggl({\s x_{n+1}+\s x_n\over2}\biggr)^s
\to\sum_{ds}\int dtG_{ds}\biggl({d\s x\over dt}\biggr)^d\s x^s,}
is $[G_{ds}]=ML^{2-d-s}T^{d-2}$.

Consider now a graph with $E$ external legs and $V$ vertices.
Out of the $E$ legs $E_d$ contain
a velocity i.e. they are attached to a velocity term at the vertex. 
The dimension of the graph can be written as
\eqn\grdim{L^E=L^{2E}T^{1-E_d}[I][G](ML^2T^{-1})^{-V}.}
The first two factors give the dimension
of the legs by taking into account $[G_0]=L^2$ and the overall energy
conservation of the loop integral. $[I]$ and $[G]$ stand for the dimension
of the loop integral and the coupling constants. Since each vertex is
divided by $\hbar$, $[\hbar]^{-V}$ appears in the last factor. The 
overall degree of divergence of the loop integral $I$ with $L$ internal 
lines is defined by $\omega[I]=-([I][\hbar]^{-L})_T$,
where $([A])_T$ is the time dimension of $A$. The removal of an $\hbar$ 
at each line is necessary to isolate the contribution of the 
energy depending factors. We find
\eqn\dimi{[I]=M^{-L}T^{-1-{E\over2}+E_d-\sum_v\omega_v},}
where the last contribution in the exponent is a sum over the vertices $v$ and
$\omega_v={d_v-s_v\over2}-1$. Thus the overall degree of divergence is
\eqn\ovdgdf{\omega[I]=1+{E\over2}-E_d+\sum_v\omega_v.}
As a quick check recall the result of the usual power counting for field 
theories in D-dimensions,
\eqn\ftdgd{\omega[I]=D+{2-D\over2}E-E_d-\sum_v[G_v]_E,}
where $[G_v]_E=D+s_v-{D\over2}(s_v+d_v)$,
is the energy dimension of the coupling constants
corresponding to the vertex $v$. Naturally \ftdgd\ reduces to \ovdgdf\
for $D=1$. For $D<2$ the graphs with more external legs are more
divergent in the ultraviolet. 
We call the vertex $v$ renormalizable or non-renormalizable for
$\omega_v<0$ or $\omega_v>0$, respectively. Even if all vertices are 
renormalizable there are divergent graphs, such as a mass insertion, 
${\delta m\over2}<{d\s x\over dt}^2>$, c.f. \ldivpr. 

It is instructive to follow what happens when we add an inhomogeneous mass term, 
$\hf M(\s x)({d\s x\over dt})^2$, to the lagrangian.
Assume the polynomial form, $M(\s x)=\sum_nM_n\s x^n$, for the
mass correction. The vertex with
the coupling constant $M_n$ has $n+2$ legs, out of those are two with
velocity coupling. The closing of the latter legs induces one-loop 
linear divergences. In order to have finite transition amplitudes we need
the counterterm 
\eqn\countim{\delta M=\sum_nM_n\s x^n{3i\hbar\over m\dt}
={3i\hbar\over2\dt m}M(\s x),}
according to \ldivpr. It will be shown below that this counterterm generates 
the one-loop effects of the integral measure of the path integral
which corresponds to the particle with kinetic energy 
$\hf(m+M(\s x))({d\s x\over dt})^2$.
The higher loop contributions will be resummed by the help of the 
infinitesimal renormalization group equations.

We close this Section by some remarks about Quantum Field Theory.
The divergence in \brownian\ is the same as predicted by the superficial degree 
of divergence of the loop integral \ldivpr. In other words, the simple 
observation that each contribution in the exponent \pathfd\ is O(1) is
in agreement with the divergences predicted by power counting. Thus
one can distinguish two kind of divergences in Quantum Field Theory.
One which is $O(\hbar)$ and another which is diverging with the number
of degrees of freedom. The former is common with Quantum Mechanics.
  
The naive power counting argument indicates that the typical field 
configurations are nowhere continuous in Quantum
Field Theory. In fact, consider the D-dimensional lattice regulated free
massless theory which is defined by the path integral
\eqn\latrnc{\int D[\phi]e^{-a^{D-2}\hf\sum_{x,\mu}(\nabla_\mu\phi_x)^2}
=\int D[\Phi]e^{-\hf\sum_{x,\mu}(\nabla_\mu\Phi_x)^2},}
where $\nabla_\mu\phi_x=\phi_{x+\hat\mu}-\phi_x$ and
the dimensionless variable, $\Phi=a^{2-D\over2}\phi$, has been
introduced. Each contribution to the action is O(1) so we have
$\nabla_\mu\Phi_x=O(1)$ and $\nabla_\mu\phi_x=O(a^{D-2\over2})$.
This discontinuity is again in agreement with the perturbative 
power counting argument in the momentum space. It can be 
obtained in a rigorous manner as well, without relying the perturbation
expansion \ref\glimmr. The nowhere continuity may lead to unexpected
consequences in the topological structure of the renormalized theory.
Apart of some isolated examples, \ref\topdefr,
it is not known how to deal with these more violent 
singularities of Quantum Field Theory in a non-perturbative manner.

In relativistic 
Quantum Field Theory the dimension of the coupling constant determines
its tree level behavior under the renormalization group,
\eqn\qftrg{k{\partial\over\partial k}g(k)=-[g]_Eg+O(\hbar),}
where $k$ is the cut-off and $[g]$ is the energy dimension of the coupling
constant $g$. Thus the irrelevant coupling constants are non-renormalizable.
The inverse time dimension of the coupling constants in Quantum Mechanics,
$-([G_{ds}])_T=2-d$, 
is different from $[G_{ds}]_E$. This is because the quantum field has 
the energy dimension $[\phi]={D-2\over2}$ while $([\s x])_T=0$.
It is $[G]_E$ or $-[G]_T$ which determines the scaling properties of the
coupling constants ? One could guess that the blocking in space or time
generates the scaling exponents $[G]_E$ or $-[G]_T$, respectively. We shall
verify this guess explicitly as far as the blocking in time is concerned.
\bigskip
\centerline{\bf 4. RUNNING COUPLING CONSTANTS} 
\medskip
\nobreak
\xdef\secsym{4.}\global\meqno = 1
\medskip
The quantity of central importance in Quantum Mechanics is the 
matrix element of the time evolution operator what we parametrize as
\eqn\prop{<\s x|U(t)|\s y>=e^{{i\over\hbar}S(\s x,\s y;t)}.}
Information about the hamiltonian can directly be collected by 
measuring the spectrum and the properties of the stationary states.
Another source of information is the investigation of the 
transition rates or scattering phenomena.
This latter refers directly to the function $S(\s x,\s y,t)$.

In the spirit of the gradient expansion we write the renormalized lagrangian as
\eqn\lgs{S(\s x,\s y;t)=\sum_{n=0}^\infty
s^{j_1,\cdots,j_n}(\xy;t)(\dx)^{j_1}\cdots(\dx)^{j_n}.}
The generalized potentials, $s^{j_1,\cdots,j_n}(\xy;t)$ have
simple form in the limit $t\to0$. In fact,
in this limit we find
$s=-tV({\xy\over2})$, $\s s=\s A({\xy\over2})$
and $s^{jk}={m\over2\dt}\delta^{jk}$, for the hamiltonian \hammd.
The time dependent running coupling constant can be introduced 
by the help of the Taylor expansion,
\eqn\rcte{s^{j_1,\cdots,j_n}(\xy;t)=\sum_{m=0}^\infty
s^{j_1,\cdots,j_n}_{k_1,\cdots,k_m}(t)(\xy)^{k_1}\cdots(\xy)^{k_m}.}
The evolution in $t$ is described by the renormalization
group equations,
\eqn\rgtd{S(\s x,\s y;t_1+t_2)=-i\hbar\ln\biggl\{
\int d\s ze^{{i\over\hbar}S(\s x,\s z;t_1)}e^{{i\over\hbar}S(\s z,\s y;t_2)}
\biggr\}.}
Note that the Planck constant, $\hbar$, is kept fixed during the
blocking, \rgtd.

In the operator formalism of Quantum Mechanics where the eigenstates and
the spectrum of the hamiltonian play a central role those measurements
are made which determine the long time behavior of the system.
The experiment which is more relevant in determining $S(\s x,\s y;t)$
is when the system is observed at the time $t_n=n\dt$, as it were
enlighten by a stroboscope. The renormalized trajectory, the 
solution of the evolution equation, \rgtd,  describes the way
the observed parameters change with the stroboscope frequency.
If the frequency of the measurements is high and no ultraviolet
divergences arise then the cut-off lagrangian approaches its 
classical form. Furthermore, the Schr\"odinger equation yields the
Hamilton-Jacobi equation with an imaginary $O(\hbar)$ drift term. 
It was this observation of Dirac, \ref\diracr, which oriented Feynman's 
attention to the path integrals.

One should bear in mind that the 
importance of the running coupling constants and the renormalization 
group is in principle independent of the eventual divergences in the theory. 
As an example we now compute the the running mass and frequency of the 
harmonic oscillator, $H={\s p^2\over2m}+{m\omega\over2}\s x^2$.
The path integral is Gaussian and we have
\eqn\hoprop{S(\s x,\s y;t)=-i\hbar\ln<\s x|e^{-i{t\over\hbar}H}|\s y>
=\hf(\s x^2+\s y^2)m\omega{\rm ctg}t\omega
-{m\omega\over\sin\omega t}\s x\s y.}

The running mass, $m(t)$, and frequency, $\omega(t)$, defined by
the parametrization
\eqn\runk{S(\s x,\s y;t)={m(t)\over2t}(\s x-\s y)^2
-{tm(t)\omega^2(t)\over2}\biggl({\s x+\s y\over2}\biggr)^2,}
are
\eqn\runmk{m(t)=\hf m(0)t\omega(0){\rm ctg}{t\omega(0)\over2},}
\eqn\runok{\omega^2(t)={4\over t^2}{\rm tg}^2{t\omega(0)\over2}.}
When the probability distribution of the coordinates is measured subsequently 
with the time difference $\dt$ in the presence of some interference then one 
can reconstruct both the real and the imaginary part of $S(\s x,\s y;\dt)$. 
In the case of the harmonic oscillator $S(\s x,\s y;\dt)$ is real and $\hbar$
independent. The system appears at these measurements as if it had the mass 
$m(\dt)$ and frequency $\omega(\dt)$. Predictions for further measurements
which are performed at integer multiple of the time $\dt$ can be obtained 
by the help of the bare theory based on \runk. 

Note that the running coupling constants are non-monotonic functions of the
observation time with singularities, $\omega(nT)=m((n+\hf)T)=0$, and 
$m(nT)=\omega((n+\hf)T)=\infty$, where $T={2\pi\over\omega(0)}$. This is the
result of making the blocking in real instead of the imaginary time.
It is easy to understand $m(nT)=0$. In fact, the propagator,
\eqn\propsp{<\s x|e^{-i{t\over\hbar}H}|\s y>=
\sum_n\psi_n(\s x)e^{-i(n+\hf)\omega_0t}\psi^*_n(\s y),}
where $H\psi_n=\hbar\omega(0)\bigl(n+\hf\bigr)\psi_n$, gives 
\eqn\propsp{<\s x|e^{-i{nT\over\hbar}H}|\s y>=(-1)^n\delta(\s x-\s y).}
The periodic self focusing is a characteristic of systems
with equidistant energy spectrum. It is reproduced by the divergences
of the effective mass because the large mass makes the coefficient of
$(\s x-\s y)^2$ large in the action which suppresses the
propagation over finite distances. The vanishing of $tm(t)\omega^2(t)$
at $t=nT$ is the immediate consequence of the fact that the
potential term is negligible compared to the kinetic energy for short
time and that the propagator is periodic in time.
The diverging potential at half period requires the initial and the
final points should have zero mean, $\s x+\s y=0$. This is the result of 
$\sum_n(-1)^n\psi_n(\s x)\psi^*_n(\s y)=\delta(\s x+\s y)$.
\bigskip
\centerline{\bf 5. DECIMATION}
\medskip
\nobreak
\xdef\secsym{5.}\global\meqno = 1
\medskip
The blocking in time gives the time dependence for $S(\s x,\s y;t)$,
\eqn\idof{e^{{i\over\hbar}S(\s x,\s y;t_1+t_2)}=\int d\s z
e^{{i\over\hbar}S(\s x,\s z;t_1)}e^{{i\over\hbar}S(\s z,\s y;t_2)}.}
In the limit $t_2=\dt\to0$ then \idof\ reduces to the Schr\"odinger equation.
For $t_1=t_2$ it gives the evolution of the renormalized action under the 
doubling of the observational time. We investigate first the scaling properties
of the coupling constants under the latter renormalization group
transformation.
\bigskip
\centerline{\bf 5a. Blocking transformation.}
\nobreak
\medskip
We use the parametrization
\eqn\sha{S(\s x,\s y;t)={m\over2t}(\dx)^2-U(\s x,\s y;t),}
where $U(\s x,\s y;t)$ will be treated as perturbation since the free action
with $U=0$ will turn out to be a fixed point.
The blocking transformation, corresponds to the choice $t_1=t_2=t$
in \idof. Note that the `nearest neighbor coupling' structure which is 
preserved during the decimation, $t\to t'=2t$, is characteristic of the 
one dimensional systems. In higher dimensions the non-local terms are 
unavoidably generated by the blocking procedure.

The contribution of the kinetic energy can be written in the exponent as
\eqn\sumkin{{m\over2t}\biggl((\s x-\s z)^2+(\s z-\s y)^2\biggr)=
{m\over t}\biggl(\s z-{\xy\over2}\biggr)^2+{m\over4t}
(\s x-\s y)^2.}
Taking the limit $t\approx0$ we expand the exponent around $\s z=\hf(\xy)$,
\eqn\expbl{
{m\over t}\s v^2+{m\over4t}(\s x-\s y)^2
-U_1-U_2-\s v\partial(U_1+U_2)-\hf v_iv_j\partial_i\partial_j(U_1+U_2),}
where $\s v=\s z-\hf(\xy)$, $U_1=U\bigl({\xy\over2},\s y\bigr)$ and
$U_2=U\bigl(\s x,{\xy\over2}\bigr)$. The gradient always acts on the 
variable which is set to $\hf(\xy)$. By truncating the resulting integral 
at the Gaussian level one finds 
\eqn\saddlsh{\int d\s ve^{-{i\over2}\s vA\s v+i\s B\s v+iC}={(-2\pi i)^{3/2}
\over\sqrt{{\rm det}A}}e^{iC+{i\over2}\s BA^{-1}\s B},}
where
\eqn\amatr{\cases{\eqalign{
\hbar A_{jk}&=-\delta_{jk}{2m\over t}+\partial_j\partial_kU_+\cr
\hbar B_j&=-\partial_jU_+\cr
\hbar C&={m\over4t}(\dx)^2-U_+,}}}
and $U_\pm=U_1\pm U_2$. The exponential prefactor can be written as
\eqn\deto{\ln{1\over\sqrt{{\rm det}A}}=-\hf\tr\ln A
={t\over4m}\bx U_++O\bigl({t^2\over m^2}\bigr).}
Finally we obtain the blocking transformation, 
$S\to{m\over2t'}(\dx)^2-U'(\s x,\s y;t')$, with
\eqn\bltru{U'=U_+
+{t'\over8m}\biggl((\partial U_+)^2-\hbar\bx U_+\biggr)+O(t^2).}
The $\hbar$ independent terms on the right hand side represent classical 
mechanics, they result from the minimization of the action. The third, 
$O(\hbar)$ term comes from the pre-exponential factor. 
For $t=O(m/\bx U_+)$ the the matrix 
$A_{jk}$ may become singular and the terms $O(t^2)$ ignored above become 
essential and generate singularities in the running coupling constants,
c.f. \runmk-\runok. This phenomenon
is similar to the emergence of the focal points of classical mechanics
\schulmanr.

It is useful to parametrize $U(\s x,\s y)$ in terms of $\s r=\dx$ and
$\s R={\xy\over2}$. To obtain $U_1$ we need the shift
$\s x\to{\xy\over2}$ which amounts to the replacement 
\eqn\she{\cases{\eqalign{\s r&\to{\xy\over2}-\s y={\s r\over2}\cr
\s R&\to\hf\bigl({\xy\over2}+\s y\bigr)=\s R-{\s r\over4}.}}}
For $U_2$ we have $\s y\to{\xy\over2}$,
\eqn\shk{\cases{\eqalign{\s r&\to\s x-{\xy\over2}={\s r\over2}\cr
\s R&\to\hf\bigl(\s x+{\xy\over2}\bigr)=\s R+{\s r\over4}.}}}
We introduce the notation $U(\s x,\s y)=\ur(\s R,\s r)$ which yields
$U_1=\ur(\s R-{\s r\over4},{\s r\over2})=\ur_1(\s R,\s r)$ and
$U_2=\ur(\s R+{\s r\over4},{\s r\over2})=\ur_2(\s R,\s r)$.
Since $\partial U_1=2\partial_x U_1$ and $\partial U_2=2\partial_y U_2$,
$\partial U_1=\bigl(\hf\partial_R+\partial_r\bigr)\ur_1$ and
$\partial U_2=\bigl(\hf\partial_R-\partial_r\bigr)\ur_2$ we have 
$\partial U_\pm=\partial_R\ur_\pm+2\partial_r\ur_\mp$ and
$\partial^2U_\pm=(\partial_R^2+4\partial^2_r)\ur_\pm
+4\partial_R\partial_r\ur_\mp$, with $\ur_\pm=\ur_1\pm\ur_2$. 
The blocking equation finally takes the form
\eqn\bltrur{\ur\to\ur_+
+{t'\over8m}\biggl\{(\partial_R\ur_++2\partial_r\ur_-)^2
-\hbar\bigl[\bigl(\partial^2_R+4\partial_r^2\bigr)\ur_+
+4\partial_R\partial_r\ur_-\bigr]\biggl\}+O(t^2).}

In order to find the scaling exponents one linearizes the blocking
relations. The linearized version of \bltrur\ contains a part of the 
classical contributions and the logarithm of the pre-exponential factor. 
The tree level term $\s BA^{-1}\s B$ in \saddlsh\ drops out leaving behind
\eqn\linbl{U\to U_+-{t'\hbar\over8m}\bx U_+.}
Notice that the U.V cut-off, $t$, appears explicitly in the blocking
relations due to the time dimension of $\hbar$. One might take this into
account by introducing the running Planck's constant, $\hbar(t)=t\hbar$,
in the action. Treating $\hbar(t)$ as an additional coupling
constant, the value of $\hbar$ at the current cut-off, one can formally
eliminate $t$ from the recursive relation \bltrur. But the
price of such a hidden treatment of the cut-off is that the linearized
blocking relation becomes the tree-level one, $U\to U_+$. So we keep
$\hbar$ as a constant and continue with the cut-off dependent relation \linbl\
in two typical cases below.
\bigskip
\centerline{\bf 5b. Polynomial velocity coupling.} 
\nobreak
\medskip
Consider the general polynomial coupling, 
\eqn\potegy{\ur=t\urv={1\over t}g\s r^dV(\s R)=G\s v^dV(\s R),}
in the lagrangian where $\s v={\s r\over t}$, and $G=gt^{d-1}$. 
The time dimension of the coupling constants should be given in the units 
of the cut-off so the factor $t^{-1}$ is inserted in \potegy\
to remove the time dimension of $g$ in the dimensionless exponent,
${g\over\hbar t}\s r^dV(\s R)$.

The linearized tree level contribution to the blocking is
\eqn\urrm{\eqalign{\ur\to\ur_+&={1\over t'}g2^{1-d}\s r^d
\biggl(V\bigl(\s R-{\s r\over4}\bigr)+V\bigl(\s R+{\s r\over4}\bigr)\biggr)\cr
&={1\over t'}g2^{2-d}\s r^d
\biggl(1+{1\over32}r^ir^j\partial_i\partial_j+O(\s r^4)\biggr)V(\s R).}}
Thus the coupling constant is rescaled during the blocking as
\eqn\resc{g\to g\biggl({t'\over t}\biggr)^\nu,}
where $\nu=2-d$ and contributions with higher powers of $\s r$ are generated. 
$\urv$ is irrelevant for $d>2$ and marginal or relevant when $d=0$ or $d=1,2$, 
respectively. Furthermore the coupling constant $G$ expressed in the usual 
units is renormalization group invariant on the tree level. 

The lesson is that the scalar and vector potentials and the mass are the 
tree level marginal and relevant terms. In non-relativistic Quantum Mechanics 
the scaling properties of the vertex $v$ is given by the time dimension 
rather than $\omega_v$, as in Quantum Field Theory because the
blocking in time does not influence the length scales in three-space.
Since $\omega_v$ is not proportional with $\nu$ non-renormalizability
and irrelevance are not equivalent, e.g. the vertices with $2<d<s+2$
are renormalizable and irrelevant.
\bigskip
\centerline{\bf 5c. Vector potential.} 
\nobreak
\medskip
The potential will be assumed of the form
$U(\s x,\s y;t)=\ur_A(\s R,\s r)+t\urv(\s R)$, with
$\ur_A=-\s r\s A(\s R+\eta\s r)$. 
The vector potential transforms as
\eqn\blvpl{\s A\to\hf(\s A_++\s A_-)+O(\bx\s A),}
where $\s A_\pm=\s A\bigl(\s R+\s r{2\eta\pm1\over4}\bigr)$.
The linearized renormalization group equations for $\urv$ is of the form
\eqn\bltrpl{\urv\to\hf\urv_+-{\hbar\over8m}\bx\ur_{A+},}
where
\eqn\derketv{
\bx\ur_{A+}=-\biggl[\bigl({1\over4}+\eta^2\bigr)\s r\bx+4\eta\partial\biggr]
(\s A_++\s A_-)+\bigl(\eta\s r\bx+2\partial\bigr)(\s A_+-\s A_-).}
It shows the mixing of the vector and the scalar potential,
\eqn\vesc{\urv(\s R)\to\urv(\s R)-{\hbar\eta\over m}\partial\s A+O(\bx\s A).}
Since the scalar potential receives a constant increment for each doubling of 
the cut-off, $\dt\to\dt/2$, the accumulation of \vesc\ gives a 
logarithmic divergence in the limit $\dt\to0$.	 

One would have thought that the absence of the ultraviolet
divergence for $\eta=0$ is similar to the reduction of the degree of
divergence due to gauge invariance in QED. We shall argue below that
this analogy is somewhat misleading. The Feynman graph responsible to
the $\eta$-dependent contribution in \vesc\ is actually finite and
such a reduction of the degree of divergence occurs even without gauge 
invariance. 
\bigskip
\centerline{\bf 6. INFINITESIMAL RENORMALIZATION GROUP EQUATION}
\xdef\secsym{6.}\global\meqno = 1
\medskip
In the decimation described in the previous Sections we
used the symmetrical setting, $t_1=t_2$, in \idof. In order to derive the 
infinitesimal renormalization group equation and the 
hamiltonian we take the limit ${t_1\over t_2}\to0$ and the integral
will be saturated by a saddle point which is different from those of the
decimation. The scaling properties can in principle different
in the lagrangian and the hamiltonian continuum limit.
\bigskip
\centerline{\bf 6a. Wegner-Haughton equations.} 
\nobreak
\medskip
We start with the action for infinitesimal time,
\eqn\shac{S(\s x,\s y;\dt)={m\over2\dt}(\dx)^2-U(\s x,\s y;t)-R,}
with 
\eqn\potham{U(\s x,\s y;t)=tV(\s R)-\s r\s A(\s R+\eta(\dx)).}
The renormalization group equation, \idof, is expanded at $\s y\approx\s x$,
\eqn\whe{e^{{i\over\hbar}S(\s x,\s y,t+\dt)}=
\int d\s ze^{-{i\over2}\s zA\s z+i\s B\s z+iC},}
where
\eqn\whexp{\cases{\eqalign{\hbar A_{jk}
&=-{m\over\dt}\delta_{jk}+{1\over4}\dt\partial_j\partial_kV
-\bigl(\hf-\eta\bigr)(\partial_jA_k+\partial_kA_j)
-\partial_j\partial_kS\cr
\hbar\s B&=-\hf\dt\partial V+\s A+\partial S\cr
\hbar C&=-\dt V-R+S,}}}
and the partial derivatives act on the first argument of $S=S(\s x,\s y,t)$.
The integration yields the relation
\eqn\whee{\eqalign{{i\over\hbar}S(\s x,\s y,t+\dt)
&={3\over2}\ln(2\pi)-\hf\tr\ln iA+{i\over2}\s BA^{-1}\s B+iC\cr
&={3\over2}\ln{2\pi i\hbar\dt\over m}-{i\over\hbar}R
-{i\dt\over\hbar}V-{\dt(1-2\eta)\over2m}\partial\s A\cr
&-{i\dt\over2m\hbar}(\s A+\partial S)^2-{\dt\over2m}\bx S+{i\over\hbar}S,}}
which can be written as a differential equation,
\eqn\wheed{\partial_tS(\s x,\s y,t)=
-\biggl({1\over2m}{\cal P}^2+V(\s x)\biggr)+{i\hbar\over2m}\partial{\cal P}
-{i\hbar\eta\over m}\partial\s A(\s x),}
with
\eqn\mom{{\cal P}=\s A(\s x)+\partial_xS(\s x,\s y,t),}
when the normalization 
\eqn\norm{R={3\hbar\over2i}\ln{m\over2\pi i\hbar\dt}}
is used.

$e^{{i\over\hbar}S(\s x,\s y,t)}$ is the wave function, $\psi(\s x,t)$, 
which satisfies the initial condition $\psi(\s x,0)=\delta(\s x-\s y)$ and
our equation \wheed\ is equivalent with the Schr\"odinger equation. 
The tree level, $O(\hbar^0)$, evolution is given by the Hamilton-Jacobi 
equation and the quantum corrections amount to
an imaginary diffusion term. It is amusing to note that our result is
equivalent with the Wegner-Haughton renormalization group equations
\ref\wegnerr\ except that it gives the time rather than the energy dependence
of the renormalized action. This can easily be seen by means of the
functional Schr\"odinger representation for Quantum Field Theory.
\bigskip
\centerline{\bf 6b. Ultraviolet scaling}
\nobreak
\medskip
As an application of the renormalization group equation we deduce
the ultraviolet scale dependence in the presence of an
external magnetic field. We take $\eta=0$ and assume the form
\eqn\trajas{S(\s x,\s y;t)=R(\s x)t^{-1}+
\sum_{\ell=0}^\infty\sum_{k=0}^\infty  S_{k\ell}(\s x)t^k\ln^\ell{t\over t_0},}
where the dependence on the initial point, $\s y$, is suppressed.
The logarithmic terms are supposed to arise in the perturbation expansion 
around the quadratic part of the action and the substraction scale is
$t_0^{-1}=O(\partial\times A/m)$.
By substituting \trajas\ into the renormalization group equation we obtain
\eqn\dlongt{\eqalign{&-Rt^{-2}+\sum_{\ell=0}^\infty\sum_{k=0}^\infty 
\biggl((k+1)S_{k+1,\ell}+(\ell+1)S_{k+1,\ell+1}\biggr)t^k\ln^\ell{t\over t_0}\cr
&={i\hbar\over2m}\biggl(\bx Rt^{-1}+\sum_{\ell=0}^\infty\sum_{k=0}^\infty 
\bx S_{k,\ell}t^k\ln^\ell{t\over t_0}\biggr)\cr
&-{1\over2m}\biggl(\partial Rt^{-1}+
\sum_{\ell=0}^\infty\sum_{k=0}^\infty\partial S_{k,\ell}t^k\ln^\ell{t\over t_0}
+\s A\biggr)^2-V-{i\hbar\over2m}\s\partial\s A}}

We now identify the time dependence on both sides,
\eqn\retk{\eqalign{{\bf O}(t^k\ln^\ell t):\  \ 
&m(k+1)S_{k+1,\ell}+m(\ell+1)S_{k+1,\ell+1}
={i\hbar\over2}\bx S_{k,\ell}-\s A\s\partial S_{k,\ell}
-\s\partial R\s\partial S_{k+1,\ell}\cr
&-\hf\sum_{n=0}^k\sum_{m=0}^{\ell}\s\partial S_{n,m}\s\partial S_{k-n,\ell-m}
-\delta_{k,0}\delta_{\ell,0}
\biggl(\hf\s A^2+mV+{i\hbar\over2}\s\partial\s A\biggr),}}
\eqn\retmln{{\bf O}(t^{-1}\ln^\ell t):\  \ m(\ell+1)S_{0,\ell+1}=
-\s\partial R\s\partial S_{0,\ell},}
\eqn\retm{{\bf O}(t^{-1}):\  \ mS_{0,1}={i\hbar\over2}\bx R
-\s\partial R\s\partial S_{0,0}-\s A\s\partial R,}
\eqn\tmk{{\bf O}(t^{-2}):\  \ -R=-{1\over 2m}(\s\partial R)^2.}

The last relation gives $R(\s x)=\hf m\s x^2$ for arbitrary potential. \retm\ 
yields $S_{0,1}={3i\hbar\over2}-\s x(\s\partial S_{0,0}+\s A)$. 
\retmln\ gives $S_{0,\ell}$ as 
$S_{0,\ell+1}=-{1\over\ell+1}\s x\s\partial S_{0,\ell}$, in a recusive manner. 
Finally the linearized form of \retk\
can be used to obtain $S_{k\not=0,\ell}$ in terms of
$S_{k,0}$. The undetermined nature of $S_{k,0}$ reflects the overcompleteness
of \trajas.

Suppose that we can neglect the terms with $k\not=0$ in the ultraviolet 
limit, $t\to0$. The remaining contributions sum up to a power divergence 
as usual,
\eqn\uvsum{\eqalign{
\sum_{\ell=0}^\infty S_{0,\ell}(\s x)\ln^\ell{t\over t_0}
&=S_{0,0}+\ln{t\over t_0}\biggl({3i\hbar\over2}
-\s x(\s\partial S_{0,0}+\s A)\biggr)\cr
&+{1\over\s x\s\partial}\sum_{\ell=2}^\infty
{1\over\ell!}\biggl(-\ln{t\over t_0}\ \s x\s\partial\biggr)^\ell
\s x(\s\partial S_{0,0}+\s A)\cr
&=S_{0,0}+{3i\hbar\over2}\ln{t\over t_0}+{1\over\s x\s\partial}\biggl[
\biggl({t\over t_0}\biggr)^{-\s x\s\partial}-1\biggr]
\s x(\s\partial S_{0,0}+\s A),}}
in terms of the initial condition for the renormalized trajectory, 
$S_{0,0}(\s x)$. It remains to be seen
if the power divergence of \uvsum\ shows up in the exact solution.
\bigskip
\centerline{\bf 6c. Hamiltonian}
\nobreak
\medskip
The hamiltonian corresponding to \shac\
will be derived by taking the limit ${t_1\over t_2}\to0$ in
\eqn\sche{\psi(\s x,t+\dt)=\int d\s ye^{{im\over2\hbar\dt}(\dx)^2
-{i\over\hbar}U(\s x,\s y;t)-{i\over\hbar}R}\psi(\s y,t).}
The quadratic approximation to the time evolution is
\eqn\schk{\psi(\s x,t+\dt)=\int d\s ze^{-{i\over2}\s zA\s z+i\s B\s z+iC}
\biggl(1+\s z\partial+\hf\s z\s z\partial\partial\biggr)\psi(\s x,t),}
with
\eqn\schka{\cases{\eqalign{\hbar A_{jk}
&=-{m\over\dt}\delta_{jk}+{1\over4}\dt\partial_j\partial_kV
-\bigl(\hf-\eta\bigr)(\partial_jA_k+\partial_kA_j)\cr
\hbar\s B&=-\hf\dt\partial V+\s A\cr
\hbar C&=-\dt V-R.}}}
The Gaussian integral gives this time
\eqn\timev{\eqalign{\psi(\s x,t+\dt)
&=(2\pi)^{3/2}e^{-\hf\tr\ln iA+{i\over2}\s BA^{-1}\s B+iC}\cr
&\biggl(1+B_jA^{-1}_{jk}\partial_k-\hf\bigl(iA^{-1}_{jk}
-(A^{-1}\s B)_j(A^{-1}\s B)_k\bigr)\partial_j\partial_k\biggr)\psi(\s x,t)\cr
&=e^{{3\over2}\ln{2\pi i\hbar\dt\over m}-{i\over\hbar}R}\cr
&\biggl(1-{i\dt\over\hbar}V-{\dt(1-2\eta)\over2m}\partial\s A
-{i\dt\over2m\hbar}\s A^2-{\dt\over m}\s A\partial
+{i\hbar\dt\over2m}\bx\biggr)\psi(\s x,t),}}
which generates the Schr\"odinger equation with
\eqn\hamin{H={(\s p+\s A)^2\over2m}+V+{i\eta\hbar\over m}\partial\s A,}
when the normalization \norm\ is used. \hamin\ is non-hermitean for real
non-vanishing $\eta$ reflecting the inappropriate order the non-commuting
operators are multiplied in the hamiltonian in this case. The hermiticity
is recovered either for the mid-point prescription, $\eta=0$, or
for imaginary $\eta$. The latter case is an example of an effective theory 
discussed below.
\bigskip
\centerline{\bf 7. LOW ENERGY EFFECTIVE THEORIES}
\xdef\secsym{7.}\global\meqno = 1
\medskip
The hamiltonian obtained in the previous Section contains a term without
classical analogy, ${i\eta\hbar\over m}\partial\s A$, which survives
the quantum continuum limit, $\dt({\Delta x\over\dt})^2=O({\hbar\over m})$,
c.f. \linvp. Thus a vector potential with special imaginary non-mid point 
prescription in the lagrangian generates a scalar potential in the hamiltonian. 
But this is not a general rule, it is valid only for expectation values
taken for a time scale which is longer than $\dt$. If the derivation of the
hamiltonian remains valid in the limit $\dt\to0$ then this latter limitation
is unimportant. But it becomes important when the ultraviolet power 
divergences of the running coupling constants bring new terms in \timev\ 
invalidating \hamin. The generalization of this phenomenon, namely the 
transmutation of the velocity coupling of the high energy ($\dt\approx0$)
lagrangian into a scalar potential of the hamiltonian,
appears to be formally similar to the construction of effective models
in Quantum Field Theory. In fact, the lagrangian defined by the help of
the cut-off dependent parameters in the limit $\dt\to0$ describes the physics 
of all time scales. Starting with such a theory one derives a different system 
given by the hamiltonian which is valid at low energies only.

Suppose that we have a renormalizable Quantum Field 
Theory which possesses a mass scale. The operators scale according
to the ultraviolet scaling laws at the high energy side of the mass
scale and we have few relevant coupling constants to parametrize the
possible interactions which can be formulated without the reference to
a minimal length scale i.e. cut-off. The mass scale represents a crossover 
to the infrared regime where different scaling laws prevail and new 
relevant operators may appear which are non-renormalizable according to the 
classification scheme of the ultraviolet scaling laws. The continuation
of the renormalized trajectory beyond the crossover usually requires
non-perturbative methods. In order to keep the perturbative treatment 
one sets up different perturbation expansions for the two scaling regime.
This amounts to the introduction of a low energy effective theory including
the operators which are relevant in the infrared scaling regime. This
theory is then matched to the original microscopic one at the crossover. 
When we make an attempt to go above the crossover energy scale with the 
effective theory then the presence of the non-renormalizable operators in
the effective lagrangian starts to cause problems. These operators are 
irrelevant at the ultraviolet fixed point and they drive the renormalized 
trajectory away from the fixed point as we try to remove the cut-off. Thus the
renormalized trajectories of the two theories are close in the infrared
scaling regime only, the effective theory deviates from the renormalized
trajectory of the microscopic theory in the ultraviolet region.

Returning to Quantum Mechanics we take the action
\eqn\gpothamm{S(\s x,\s y;\dt)={m\over 2\dt}(\dx)^2
-\dt^{1-d}(\dx)^dV\biggl({\xy\over2}\biggr)-R}
where the potential is chosen to be renormalization group invariant,
i.e. the $\dt$ dependence of $S$ corresponds to a renormalized trajectory. 
The evolution equation is
\eqn\schgl{\psi(\s x,t+\dt)=
\int d\s ze^{-{m\over2i\hbar\dt}\s z^2
-{i\over\hbar}\dt^{1-d}V(\s x)\s z^d-{i\over\hbar}R}
\biggl(1+\s z\partial+\hf\s z\s z\partial\partial\biggr)\psi(\s x,t),}
and the resulting hamiltonian,
\eqn\hamevm{H=-{\hbar^2\over2m}\bx+{1\over({d\over2})!!2^{d/2}}
\biggl({i\hbar\over\dt m}\biggr)^{d\over2}V(\s x),}
is obtained in the $d\over2$-th order of the loop expansion. 
The following observations are in order at this point: 

(i) The hamiltonian
which appears in the differential equation should not contain the cut-off
explicitly because the original lagrangian, \gpothamm, is 
renormalization group invariant. What happens is that we expand around the 
time-dependent trajectories in the lagrangian version of the
renormalization group i.e. the decimation and there are low order loop 
corrections which generate contributions $O((\dx)^p)$ in the action with 
$0<p<d$. Since $\dx=0$
for the saddle point in identifying the hamiltonian these contributions are 
absent in the perturbation expansion of Section II and in \hamevm.
The cut-off dependence in \hamevm\ thus reflects the difference between the 
scaling laws in the lagrangian and the hamiltonian case, $t_1=t_2$ and
$t_1<<t_2$, respectively in \idof. 

(ii) Both 
\eqn\potnsa{S_{micr}(\s x,\s y;\dt)={m\over2\dt}(\dx)^2-\dt 
V\biggl({\xy\over2}\biggr),}
and
\eqn\potr{S_{eff}(\s x,\s y;\dt)={m\over2\dt}(\dx)^2-
\dt({d\over2})!!2^{d\over2}\biggl((\dx)^2{m\over i\hbar\dt}\biggr)^{d\over2}
V\biggl({\xy\over2}\biggr),}
yield the same hamiltonian, $H=-{\hbar^2\over2m}\bx+V$ according to the 
argument leading to \hamevm. This result seems reasonable since 
${(\dx)^2m\over i\hbar\dt}=1$ for small $\dt$, c.f. \brownian, \itopr. 
Is there any difference between these two systems ? 

The answer to this question lies in the scaling properties of the 
coupling constants. On the one hand, the lagrangian \potnsa\ is renormalization
group invariant so it scales appropriately and the cut-off can be removed. On
the other hand, \potr\ does not belong to a renormalized trajectory because
the potential is suppressed in the ultraviolet. For the derivation of the
hamiltonian \hamevm\ from \potr\ $\dt$ should satisfy two conditions:
On the one hand, it must be small enough to neglect the higher orders of 
the Taylor expansion of the wave function. On the other hand, 
it has to be large enough to keep the 
potential as a perturbation to the kinetic energy. 
Once the hamiltonian is obtained the time evolution obviously
follows for $t>\dt$. So the predictions of the hamiltonian 
\hamevm\ differ from those of the lagrangian \potr\ in the ultraviolet regime 
and the physical content of \potnsa\ and \potr\ agrees up to a certain energy 
level only.

(iii) The fundamental difference between the operator and the 
path integral formalism is that the path integral expressions are well defined 
for finite value of the cut-off, $\dt$. The basic quantity
of the path integral formalism is $S(\s x,\s y;\dt)$. The convergence of the 
path integral in the continuum limit, $\dt\to0$, is a rather involved 
question due to the presence of ultraviolet divergences encountered above.
The hamiltonian is obtained in the limit $\dt\to0$ from $S(\s x,\s y;\dt)$. 
In this context the Schr\"odinger equation is a renormalized 
differential equation where the cut-off is already removed.

It is worthwhile comparing this situation with Quantum Field Theory.
The low energy effective models of Quantum Field Theories contain the vertices 
which are important at long distances. Most of these vertices are not 
renormalizable and only a finite dimensional sub-class of the
effective interactions can be derived from a renormalizable theory. The
non-renormalizable vertices can not be kept in 
the effective lagrangian without the help of the ultraviolet cut-off. 
Thus the class of not necessarily renormalizable
effective theories is wider than the family of the renormalizable models. 
Similarly, certain interactions can only be described by a finite value of the 
cut-off, $\dt$, in the path integral formalism of Quantum Mechanics. When the
limit $\dt\to0$ is taken to derive the hamiltonian some of them
become inconsistent. In high energy physics the interactions which can not be 
obtained from a renormalizable model are excluded because the
ultimate laws are supposed to be valid down to arbitrary short 
distances. Non-relativistic Quantum Mechanics is not a fundamental
theory since relativistic multi-particle effects appear at high energies.
Thus the restriction of the interactions by the removability
of the cut-off is not imposed and the bare path integral formalism becomes
more general than the renormalized operator formalism. 
\bigskip
\centerline{\bf 8. CURVED SPACES}
\xdef\secsym{8.}\global\meqno = 1
\medskip
The free particle hamiltonian on a curved space is given by the Laplace 
operator,
\eqn\laplc{H=-{\hbar^2\over2m}{1\over\sqrt{g}}\partial_ig^{ij}\sqrt{g}\partial_j
=-{\hbar^2\over2m}(g^{ij}\partial_i\partial_j-g^{jk}\Gamma_{jk}^i\partial_i),}
where $g={\rm det}g_{ij}$ and $\Gamma_{jk}^i$ is the Christoffel symbol.
The distribution of
the factors $\sqrt{g}$ is not unique in classical mechanics and is given
here by the Laplace operator as a special ordering of non-commuting operators.
We address now the question whether this special ordering and the underlying
geometry can be identified in the path integral formalism by
inspecting the renormalized trajectory. 

Let us start with the classical lagrangian, 
$L=\hf m(|\s x|)({d\s x\over dt})^2$, which defines the hamiltonian 
$H=-{\hbar^2\over2m(|\s x|)}(\bx-\hf\partial\ln m(|\s x|)\partial)$. 
For simplicity we assume spherical symmetry. The corresponding path integral 
can be obtained by the standard methods, \schulmanr, 
\eqn\cspi{S(\s x,\s y,\dt)=\tilde S(\s x,\s y,\dt)-{\hbar^2\over12}R_G\dt
+i\hbar\biggl({3\over2}\ln2\pi i\hbar-\hf\ln D(\s x,\s y,\dt)\biggr).}
The action contains the higher order terms in the derivative due to the
non-differentiability of the trajectories,
\eqn\actc{\tilde S={m(R)\s r^2\over2\dt}\biggl[1
+{m'(R)\over24m(R)}\biggl({1\over R}-{m'(R)\over2m(R)}\biggr)\s r^2
+{m'(R)\over24m(R)R^2}
\biggl({m''(R)\over m'(R)}-{m'(R)\over m(R)}-{1\over R}\biggr)
(\s R\s r)^2\biggr].}
The van Vleck determinant is 
\eqn\vvd{D=\biggl({m(R)\over\dt}\biggr)^3
\biggl(1-{\s r^2m'(R)\over8Rm(R)}\biggr)^2
\biggl(1-{\s r^2m''(R)\over8m(R)}\biggr),}
and the It\^o potential, the second term on the right hand side of \cspi\
is given by the curvature,
\eqn\curva{R_G={1\over m(R)}\biggl[4{m'(R)\over Rm(R)}+2{m''(R)\over m(R)}
-{3\over2}\biggl({m'(R)\over m(R)}\biggr)^2\biggr].}
The integration is performed by the flat measure, $d\s x$.

Suppose that we start with the action
\eqn\tract{S(\s x,\s y,\dt)={1\over2\dt}m(R)\s r^2+
{1\over\dt}a(R)\s r^4+{1\over\dt}b(R)\s r^2(\s r\s R)^2.}
The renormalized trajectory should predict the evolution of the
ultraviolet divergent and the irrelevant coupling constants of the lagrangian 
because these are the parameters which need fine tuning in order to arrive 
at finite observables in the continuum limit. We have already noticed at 
eq. \countim\ that the one-loop finiteness requires the counterterm 
${3i\hbar\over2}\ln m(\s x)$ in the action. Such a counterterm reproduces 
the one-loop contribution to the first factor of the van Vleck determinant. 
The $O(\s r^2)$ pieces of \vvd\ yield ultraviolet finite
contribution and can not be determined easily by the 
renormalization group technique.

The $O(r^4)$ irrelevant pieces of the action $\tilde S$ can be reproduced
by requiring renormalization group invariance, i.e. by following the
the renormalized trajectory. The blocking of \tract\ gives
\eqn\blomix{\eqalign{S\to&{1\over2\dt}m(R)\s r^2+
{\s r^2\over64\dt}\biggl[\s r^2{m'(R)\over R}+{(\s r\s R)^2\over\s R^2}
\biggl(m''(R)-{m'(R)\over R}\biggr)\biggr]\cr
&{1\over4\dt}a(R)\s r^4+{1\over4\dt}b(R)\s r^2(\s r\s R)^2+O(\s r^5),}}
according to \urrm. The condition for $a$ and $b$ be scaling functions is
\eqn\amix{\cases{\eqalign{&{m'(R)\over64 R}+{a(R)\over4}=a(R)\cr
&{1\over64R^2}\biggl(m''(R)-{m'(R)\over R}\biggr)+{b(R)\over4}
=b(R),}}}
whose solution is
\eqn\solsc{\cases{\eqalign{a(R)&={1\over48R}m'(R)\cr 
b(R)&={1\over48R}m'(R)\biggl({m''(R)\over m'(R)}-{1\over R}\biggr),}}}
the $O(m')$ pieces of \actc. The $O(m'^2)$ part of the action which is 
quadratic in the Christoffel symbol can be generated by the two-loop 
blocking equations.

The higher loop terms of the action are determined uniquely by
the infinitesimal form of the renormalization group equation because it is
equivalent with the Schr\"odinger equation. We take the action, 
\eqn\gpothamk{S_m(\s x,\s y;\dt)
={m(\s R)\over2\dt}\s r^2-W_m(\s R)-\dt W_I(\s R)+\s r\s A(\s R).}
The corresponding expression for $\psi(\s x,t+\dt)$, \schk, contains
\eqn\schkb{\cases{\eqalign{\hbar A_{jk}
&=-{m\over\dt}\delta_{jk}
+{1\over4}\partial_j\partial_k(W_m+\dt W_I)
-\hf(\partial_jA_k+\partial_kA_j)\cr
\hbar\s B&=-\hf\partial(W_m+\dt W_I)+\s A\cr
\hbar C&=-W_m-\dt W_I,}}}
The resulting equation of motion,
\eqn\timevb{\eqalign{\psi(\s x,t+\dt)&=e^{-{i\over\hbar}W_m
+{3\over2}\ln{2\pi i\hbar\dt\over m}}
\biggl(1+{i\hbar\dt\over2m}\bx-{i\dt\over\hbar}W_I+{\dt\over8m}\bx W_m\cr
&-{\dt\over2m}\partial\s A-{i\dt\over2\hbar m}
\bigl(\s A-\hf\partial W_m\bigr)^2
-{\dt\over m}\bigl(\s A-\hf\partial W_m\bigr)\partial\biggr)\psi(\s x,t),}}
defines the hamiltonian
\eqn\hamk{H=-{\hbar\over2m}\bx-{i\hbar\over m}
\bigl(\s A-\hf\partial W_m\bigr)\partial
+W_I-{i\hbar\over2m}\partial\s A
+{i\hbar\over8m}\bx W_m+{1\over2m}\bigl(\s A-\hf\partial W_m\bigr)^2,}
as long as $W_m(\s x)=R$, where $R$ is given by \norm.
This latter condition determines the integration measure as $d\s x m^{3/2}(x)$
up to a constant. The hamiltonian becomes 
\eqn\hamk{H=-{\hbar^2\over2m^{3/2}}\partial m^\hf\partial+V.}
with the choice $\s A=-i\hbar\partial\ln m$ and
\eqn\itopch{W_I=V+{\hbar^2\over12}R_G.}
\nobreak
\medskip
\centerline{\bf 9. QUANTUM ANOMALY AND OPERATOR ORDERING}
\xdef\secsym{9.}\global\meqno = 1
\medskip
The $\eta$ dependence has already served as a simple demonstration of the
importance of the non-differentiability of the trajectories and 
the construction of effective theories. Another interesting question this 
$\eta$ dependence raises is an analogy with the
chiral anomaly in Quantum Field Theory. 

First note that we may consider 
$\eta$ as a parameter which characterizes different regularizations since
it drops out in the classical continuum limit and 
influences the action at the cut-off scale only. But the leading
order effect of $\eta$ is more than a regularization dependence.
One can see this by recalling that the regulators are always 
irrelevant operators. The higher order contributions to the action,
$O(\eta^k)\approx\dt^k\int dt\s v^{k+1}(t)$, with $k>1$ are indeed irrelevant
but the leading order contribution, $k=1$, is marginal on the tree level.

Another important observation is that the regularization with $\eta=0$ 
preserves the U(1) gauge covariance of the transition amplitudes. 
For $\eta\not=0$ the $\eta$ dependent counterterms generate
a superficially logarithmically divergent graph in $O(\eta)$, 
${1\over\Lambda}(a\Lambda+b\Lambda\ln\Lambda)$, c.f. \linvp.
Due to the absence of a scale parameter in the background trajectory
we expand around, $x_{cl}(t)=0$, the logarithmic contribution is
absent, $b=0$ and the graph is finite, ${1\over\Lambda}a\Lambda$. 
Since the higher powers of $\eta$ multiply irrelevant operators 
the $\eta$ dependence comes from the leading order, $O(\eta)$, only. 

This situation is to be compared with the chiral anomaly where the
symmetry breaking pattern for the vector and axial charges is 
determined by the choice of the cut-off, i.e. the regularization.
The anomalous contribution is usually given by superficially
divergent graphs which turn out to be finite. Furthermore the effect
is genuinely one-loop, radiative corrections are absent.

The connection between the $\eta$ dependence and the operator ordering problem
leads to another analogy. Consider a gauge 
theory in lattice regularization where the
gradients are replaced by finite differences and exponantialized link 
variables and and nontrivial integration measures are introduced to preserve 
the gauge symmetry. One finds non-polynomial interactions and complicated
lattice propagators in perturbation expansion but the deviation from
the continuum propagators and vertices is suppressed by the lattice spacing. 
May we send the lattice spacing to zero before carrying out the
loop integration and thereby arguing that the lattice artifacts drop out
in the perturbative continuum limit ? The answer is not trivial because the
genuine lattice vertices are non-renormalizable and may generate new
ultraviolet divergences which can compensate the tree level suppression
of the vertices. Nevertheless one can prove that for certain class
of models the lattice perturbation expansion converges to the result
of the continuum regularization and the lattice artifacts are suppressed
in the continuum limit \rieszpcr. The proof starts with the properly
substracted theory where all loop integrals are made finite by the help
of the counterterms. If the convergence 
of the loop integrals is uniform then the continuum limit can be taken
before the loop integration. In other words, any vertex which is vanishing in 
the classical continuum limit gives no contribution in the quantum continuum 
limit either. 
The condition used in the proof is that all loop integral
is uniformly convergent. This holds for finite integrals with negative overall
degree of divergences. The loop integrals which have non-negative overall 
degree of divergences but happened to be finite may not converge uniformly.

We know three exceptions when classically suppressed vertices play role
in the renormalized quantum theory. 
(i) The loop integrals which are finite so require no divergent counterterm but 
remain sensitive to the presence of the cut-off due to their nonuniform
convergence are the source of the anomalies in Quantum Field Theories.
(ii) The fermion species doubling is reduced on the lattice by adding
a piece to the action which is vanishing in the classical continuum limit.
Nevertheless they remove particle modes in the quantum theory.
(iii) Operator ordering problem, i.e. $\eta$ dependence in Quantum Mechanics.

Note that the $\eta$ dependence becomes divergent when the background 
trajectory is chosen to be non-constant, c.f.\vesc. This raises some 
questions concerning our understanding of these peculiar cut-off effects
because they are usually analyzed by means of the perturbation expansion 
around the homogeneous vacuum.
\nobreak
\medskip
\centerline{\bf 10. SUMMARY}
\xdef\secsym{10.}\global\meqno = 1
\medskip
The method of renormalization group was applied to non-relativistic
Quantum Mechanics in this paper. Our motivation was based on the
guess that the non-differential, fractal nature of the trajectories 
in the path integral may generate strong effective velocity couplings at high
energy beyond the prediction of the semiclassical limit.

The power counting was developed by viewing Quantum Mechanics as a 0+1
dimensional Quantum Field Theory. A vertex with $d$ and $s$ legs attached
to velocity and coordinate, respectively, was found non-renormalizable if
$\omega={d-s\over2}-1>0$. In Quantum Field Theory the non-renormalizable
vertices are irrelevant, i.e. their energy dimension determines their
tree level scaling properties. We met an interesting question here: 
The inverse time
dimension of a vertex is different in the 0+1 dimensional Quantum Field Theory
and in Quantum Mechanics since the field and the coordinate have different
dimensions. This reflects the difference of the scaling laws one finds when the
blocking is made in the space or in the time for a non-relativistically
invariant system. To find the proper scaling laws we computed the evolution of 
the coupling constants in Quantum Mechanics by decimation, i.e. under the 
change $\dt\to2\dt$
of the cut-off in the path integral. It was found that the inverse
time dimension, $\nu=2-d$, gives the scaling exponent of the coupling
constants in the tree level. The difference between $\omega$ and $\nu$
explains of the existence of renormalizable 
and in the same time irrelevant terms in Quantum Mechanics, such as
$\s v^d\s x^s$, with $2<d<s+2$.

The relevant terms of the action are the scalar potential 
and the coupling to a gauge field, with $\nu=2$ and 1, respectively.
The mass term is marginal, $\nu=0$. This result explains the lack of need
for any other terms beyond the usual ones to characterizes
the continuum of the ultraviolet fixed points. The relevant and
marginal terms of the action have finite limit as $\dt\to0$ apart
of the case of a gauge potential with non-midpoint prescription. One
may add other renormalizable but irrelevant terms to the action without
changing the low energy behavior of the theory. 

The blocking transformation was performed in real time. The nontrivial
superposition of the complex amplitudes leads to surprising periodic
effects for harmonic systems. It was verified in the case of the harmonic
oscillator that the running mass and frequency develop singularities
at finite values of the cut-off as the result of a self focusing. These
singularities survive in any finite order of the perturbation expansion
and suggest the need of going beyond the perturbative approaches in 
describing the real time dependence of the propagation.

The infinitesimal form of the renormalization group equation, the Schr\"odinger
equation for the logarithm of the wave function, was derived as well.
It displays different scaling behavior than the blocking. This
circumstance was exploited to construct low energy effective theories.

It was pointed out that the hamiltonian operator formalism which
corresponds to the renormalized limit $\dt\to0$ is formally more restrictive
than the path integral representation. This is because the cut-off
is small but finite in the latter which can in this manner include 
non-renormalizable interactions. In other words, there are
quantum systems which admit regulated path integral description
but have no finite hamiltonian. But it is not clear whether there are
real physical systems which would correspond to these mathematical
possibilities. To decide this question
one has to determine the scale where the non-renormalizable 
coupling constants start to grow as we increase the energy. If this scale 
happens to be below the onset of the relativistic effects then these
systems could in principle be found in Nature.

Another remark concerning the ultraviolet divergences is that their 
regularization is apparently a highly nontrivial matter in 
path integral representation of the non-relativistic Quantum Mechanics. 
The Quantum Field Theories can be
regulated by omitting modes with high enough energy. The resulting
finite hamiltonian can be expressed in terms of the canonical variables.
In non-relativistic Quantum Mechanics with one degree of freedom 
there is no canonical pair of variables to neglect when the high frequency
modes are omitted. It is not obvious how
can we split the only physical degree of freedom of the quantum system
to render the theory finite by retaining the canonical structure.

Fortunately we seem to have no problem in the canonical operator formalism
where the hamiltonian must remain finite as the cut-off is removed.
This is because almost all coupling constants which can be the source of divergences 
in the hamiltonian are irrelevant. The only tree-level marginal
ultraviolet divergent vertex of the lagrangian is related 
to the non-midpoint prescription of a vector potential but it 
generates finite hamiltonian. Thus no regularization is needed
for the canonical operator formalism of Quantum Mechanics. 
Though this reassuring result is based on the tree level scaling laws.
A troublesome divergence was found in Section 6b which resulted from a 
partial resummation of the perturbation
expansion. Since there is no known exact solution for the propagator
in a general inhomogeneous magnetic field we can not decide whether
this divergence is really present in the operator formalism.

The fractal structure of the trajectories of the path integral amplifies
the effects of the velocity dependent terms in the action. In particular,
a term $\dt\int dt\s v^2$ gives finite contribution since
$\dt^2\s v^2=d\s x^2=O(\dt)$. Thus there are classically vanishing
velocity dependent terms in the action which lead to finite effects
in the continuum limit, $\dt\to0$. This is a characteristic feature of
the quantum propagation. Had the typical trajectory of the path integral
been differentiable the classically vanishing terms of the action would have
been irrelevant. Thus certain pieces of the action which are important in
the quantum dynamics have no room in classical geometry which is based on 
smooth curves, surfaces, etc. But the effects of these terms can be summarized 
in a generalized It\^o-potential which appears in the hamiltonian in higher
order of $\hbar$. This contribution together with the integral measure
which corresponds to the geometry of the system can be extracted from the
renormalized trajectory.

Finally we list some open questions or problems we encountered
during this work.

(i) We considered decimation in time. The time and the length
dimensions are not related in the non-relativistic systems so
one expects that the blocking in space and time will reveal different aspects
of the quantum dynamics. Blocking in space seems rather unusual since
there are no real degrees of freedom to eliminate as in statistical
mechanics or field theory. It leads to the elimination of
variables in finite difference equation rather than the integration over
variables of the path integral. Is it possible to cast the thinning of 
a single degree of freedom, i.e. the adjustment of the changing resolution 
in space into the same formalism in the first and second quantized systems ?

(ii) What are the scaling functions in Quantum Mechanics ?

(iii) Can one generalize Ehrenfest theorem for non-renormalizable
Quantum Mechanics where the classical limit misses certain pieces of the
lagrangian ?

(iv) Can there be an infrared fixed point in Quantum Mechanics
with nontrivial relevant or marginal terms ? Do operators which are
renormalizable and irrelevant in the ultraviolet scaling regime 
play any role in the dynamics ?

Two further questions concerning field theory:

(v) The decimation in real time leads to an interesting, nontrivial 
renormalized trajectory in the case of a harmonic oscillator. What is
the corresponding result for Quantum Field Theories where the
singularities of the running parameters should be distributed
continuously in time ?

(vi) The amplification of the velocity dependent interactions due to
the non-differentiable nature of the trajectories can be taken into
account in non-relativistic Quantum Mechanics by a potential in the 
hamiltonian. What is the analogous situation Quantum Field Theories
where the typical field configurations of the path integral are nowhere
continuous ? 
\bigskip
\centerline{\bf ACKNOWLEDGMENT}
\medskip
\nobreak
\par The author thanks Dominique Boose, Vincenzo Branchina, Kenneth Johnson
and Helmut Kroger for valuable discussions.
\bigskip
\centerline{\bf REFERENCES}
\medskip
\nobreak
\def\pr{{\it Phys. Rev.\ }}
\def\pre{{\it Phys. Rep.\ }}
\def\re{{\it Rev. Mod. Phys.\ }}

\def\pl{{\it Phys. Lett.\ }}
\def\cmp{{\it Comm.Math. Phys.\ }}
\def\jmp{{\it J. Math. Phys.}}
\def\ptp{{\it Progr. Theor. Phys.\ }}

\def\wilsrg{K. G. Wilson, \pr {\bf B4} (1971) 3174;
K. G. Wilson and J. Kogut, \pre {\bf 12 C} (1974) 75 ;
K. G. Wilson, \re {\bf 47} (1975) 773. }
\def\gosd{P. Gosdzinsky and R. Tarrach, {\it Amer. J. Phys.} {\bf 59} (1991) 
70.}
\def\gup{K.S. Gupta and S.G. Rajeev, \pr {\bf D48}(1993) 5940,
hep-th/9305052}
\def\manuel{C. Manuel and R. Tarrach, \pl {\bf B328} (1994) 113,
hep-th/9309013.}
\def\river{L. J. Boya and A. Rivero, "Renormalization in 1-D Quantum Mechanics: 
Contact Interaction", preprint DFTUZ 9413, University of Zaragoza, 
hep-th/9411081}

\def\nieuw{K. Skenderis, P. van Nieuwenhuizen, "On the Hamiltonian
Approach and Path Integration for a Point Particle Minimally
Coupled to Electromagnetism", ITP-SB-93-86, hep-th/9401024.}
\def\orderp{Masa-aki Sato \ptp {\bf 58} (1977) 1262.}
\def\schulman{L.S. Schulman,
{\it Techinques and Applications of Path Integrations}, Wiley, New York, 1981.}
\def\glimm{J. Glimm and A. Jaffe, {\it Quantum Physics A Functional Point
of View} 1987, Springer Verlag.}
\def\topdef{J. Polonyi, in the Proceedings of the Workshop on 
{\it QCD Vacuum Structure and Its Applications}, H.M. Fried and
B. Muller eds., World Scientific, 1993 pag. 3.}
\def\rieszpc{T. Reisz, \cmp {\bf 116} (1988) 81; {\bf 116} (1988) 573;
{\bf 117} (1988) 79; {\bf 117} (1988) 639.}
\def\itop{D. McLaughlin and L. S. Schulman \jmp {\bf 12} (1971) 2520.}
\def\itocal{K. It\^o, {\it Mem. Am. Math. Soc.} No. 4 (1951).}
\def\dirac{P. A. M. Dirac, {\it Physikalische Zeitschrift der Sowjetunion},
{\bf 3} (1933) No. 1.}

\def\wegner{F. J. Wegner, A. Haughton, \pr {\bf A8} (1973) 40.}
\item{\wilsrgr}\wilsrg
\item{\gosdr}\gosd
\item{\gupr}\gup
\item{\manuelr}\manuel
\item{\riverr}\river
\item{\nieuwr}\nieuw 
\item{\orderpr}\orderp
\item{\schulmanr}\schulman
\item{\itocalr}\itocal
\item{\itopr}\itop
\item{\rieszpcr}\rieszpc
\item{\glimmr}\glimm
\item{\topdefr}\topdef
\item{\diracr}\dirac
\item{\wegnerr}\wegner
\vfill
\eject
\end